\documentstyle[12pt,equation,epsfig]{article}

\newcommand{\be}{\begin{equation}}
\newcommand{\ben}{\begin{subequations}}
\newcommand{\een}{\end{subequations}}
\newcommand{\beq}{\begin{eqalignno}}
\newcommand{\eeq}{\end{eqalignno}}
\newcommand{\ee}{\end{equation}}

\newcommand{\tanb}{\tan \! \beta}
\newcommand{\tgq}{\tilde{g}-\tilde{q}}
\newcommand{\tchi}{\tilde{\chi}}

\newcommand{\ms}{M_{\rm S}}

\newcommand{\s}{\\ \vspace*{-3.5mm} }

\def\gsim{\:\raisebox{-0.5ex}{$\stackrel{\textstyle>}{\sim}$}\:}

\begin{document}  

\pagestyle{empty}
\begin{flushright}
PM/00--14\\
TUM-HEP-370-00 \\
April 2000
\end{flushright}

\vspace{1cm}

\begin{center}

{\large \bf QCD Corrections to Neutralino--Nucleon Scattering }

\vspace{1cm}

Abdelhak Djouadi$^1$ and Manuel Drees$^2$

\vspace{0.5cm}

{}$^1${\it Laboratoire de Physique Math\'ematique et Th\'eorique,
UMR5825--CNRS,\\ Universit\'e de Montpellier II, F--34095 Montpellier
Cedex 5, France.}

\vspace*{2mm}

{}$^2${\it Physik Department, Technische Universit\"at M\"unchen, \\ 
James Franck Strasse, D--85748 Garching, Germany}
\end{center}

\vspace{2cm}

\begin{abstract}

\noindent 
We calculate the dominant loop corrections from both standard and
supersymmetric QCD to the effective coupling of neutralinos to
nucleons. The potentially largest corrections come from gluino--squark
loop contributions to the Higgs boson couplings to quarks; these 
corrections also affect the leading spin--independent squark exchange
contribution. Ordinary QCD corrections to the effective coupling of
CP--even Higgs bosons to two gluons are also sizable. For large $\tanb$
values, i.e. in the region of parameter space probed by current and
near--future direct Dark--Matter search experiments, the total corrections 
can exceed a factor of three.

\end{abstract}

\newpage
\def\thefootnote{\arabic{footnote}}
\setcounter{footnote}{0}
\setcounter{page}{1}
\pagestyle{plain}

\section*{1. Introduction}
There is convincing evidence that dark (non--luminous) matter (DM)
contributes between 20 and 90\% of the critical density of the
Universe, with values around 30 to 40\% currently being favored
\cite{1}. Studies of Big Bang nucleosynthesis show that this DM cannot
be formed by baryons \cite{2}, unless they are bound in primordial
black holes. Light neutrino DM is strongly disfavored by analyses of
galaxy formation \cite{3}. One thus needs physics beyond the Standard
Model (SM) to explain the dark matter in the Universe. The probably
best motivated, and certainly most frequently studied, particle
physics candidate is the lightest neutralino $\tchi$ \cite{4}
predicted by supersymmetric extensions of the SM. In the Minimal
Supersymmetric extension of the Standard Model (MSSM) \cite{17}, its
stability against decay into ordinary particles can be guaranteed by a
symmetry called $R-$parity \cite{Fayet}, and for wide regions of the
supersymmetric parameter space one predicts a relic density in the
desired range. \s

These arguments in favor of neutralino DM have motivated several
experimental groups around the world to search for ambient relic
neutralinos. The strength of the expected signal in the two most
promising search strategies is directly proportional to the
neutralino--nucleon scattering cross section $\sigma_{\tchi N}$; these
are the search for high--energy neutrinos originating from the
annihilation of neutralinos in the center of the Sun or Earth (the
so--called ``indirect detection'') \cite{5}, and the search of the
elastic scattering of ambient neutralinos off a nucleus in a
laboratory detector (``direct search'') \cite{6}. An accurate
calculation of $\sigma_{\tchi N}$ for given model parameters is thus
essential for the interpretation of the results of these searches. \s

The matrix element for $\tchi N$ scattering, mediated by Higgs bosons, 
$Z$ boson and squark exchange diagrams, receives both
spin--dependent and spin--indepen\-dent contributions. The former are
important for neutralino capture in the Sun, but are irrelevant for
capture in the Earth, and play a subdominant role in most direct
search experiments, which employ fairly heavy nuclei (Si, Ge, I,
Na). The spin--independent contribution in turn is usually dominated
by Higgs exchange diagrams, where the Higgs bosons couple either
directly to light ($u,d,s$) quarks in the nucleon, or couple to two
gluons through a loop of heavy ($c,b,t$) quarks or squarks. In some
cases squark exchange contributions can also be important. \s

In this note we point out that these spin--independent contributions
receive large QCD corrections. The effective Higgs couplings to gluons
receive standard QCD corrections with large positive coefficients,
related to the QCD $\beta-$function. In addition, gluino--squark loop
corrections to the quark masses (or to the Yukawa couplings, for fixed
physical quark masses) can become very large for down--type quarks, 
and for the $s$ quark in particular. We
compute both kinds of corrections, and argue that corrections to other
$\tchi N$ scattering diagrams should be small. The required formulae
are given in Section 2, while Section 3 contains some numerical examples. 
In Section 4 we briefly summarize our results and draw some conclusions.

\section*{2. Formalism}

The spin--independent $\tchi N$ scattering amplitude receives
contributions from Higgs bosons \cite{7} and squark \cite{8,9} exchange
diagrams. Models with unified gaugino masses typically have squark
masses much larger than the neutralino mass, $m_{\tilde q} \gsim 5 
m_{\tchi}$, in which case the dominant contribution usually comes from 
Higgs boson exchange. Note that only scalar
Higgs couplings to neutralinos contribute in the non--relativistic
limit. In the absence of significant CP--violation in the Higgs sector, 
one therefore only has to include contributions of the two neutral
CP--even Higgs bosons. The contribution of the heavier Higgs boson
often dominates, since its couplings to down--type quarks are enhanced
if the ratio of vacuum expectation values (vev's) is large, $\tanb \gg
1$. \s

CP--even Higgs bosons can couple to nucleons in two different ways:
through an effective coupling to two gluons mediated by loops of heavy
quarks \cite{10} or squarks \cite{9}, or directly to light quarks in
the nucleon. The leading contribution to the $H^0_igg$ couplings comes
from heavy quark triangle diagrams. To leading order, they can be
described by the effective Lagrangian
\be \label{e1}
{\cal L}_{H_i^0gg}^{(Q),0} = - H_i^0 F_{\mu \nu a} F^{\mu \nu a} \frac
{\alpha_s} {12 \pi} \sum_{Q=c,b,t} \frac {c_{iQ}} {M_W} ,
\ee
where $H_i^0$ is one of the neutral CP--even Higgs bosons $h$ and $H$ 
and $F_{\mu \nu a}$ is the gluon field strength tensor, $a$ being a color
index. Note that the dimensionless coefficients $c_{iQ}$ are
independent of $m_Q$; the factor $m_Q$ in the $H_i^0 \bar{Q} Q$
coupling is canceled by a factor $1/m_Q$ from the loop integral
\cite{10}. Explicit expressions for these coefficients can e.g. be
found in Ref.~\cite{9}. The effective Lagrangian eq.~(\ref{e1}) gives rise
to $H_i^0 \bar{N} N $ couplings, $N$ being a nucleon, through hadronic
matrix elements \cite{9,10}
\be \label{e2}
\frac{\alpha_s} {4 \pi} \langle N | F_{\mu \nu a} F^{\mu \nu a} | N
\rangle = - \frac {2}{9} m_N \left( 1 - \sum_{q=u,d,s} f_{Tq} \right),
\ee
where we have introduced
\be \label{e3}
f_{Tq} = \frac {m_q} {m_N} \langle N | \bar{q} q | N \rangle.
\ee

The first point we wish to make is that the effective interaction
(\ref{e1}) is subject to large corrections from ordinary QCD. These
can most easily be computed using low--energy theorems \cite{11a}. 
The effective Higgs--gluon interaction can then be rewritten as
\be \label{e4}
{\cal L}_{H_i^0gg}^{(Q)} = - \frac{1}{4} \frac {\beta_Q(\alpha_s)} { 1
+ \gamma_Q(\alpha_s) } F_{\mu \nu a} F^{\mu \nu a} \sum_Q \frac
{c_{iQ}} {M_W}.
\ee
Here $\beta_Q$ is the contribution of heavy quark $Q$ to the QCD
$\beta-$function and $\gamma_Q$ is the anomalous dimension of
$m_Q$. Eq.(\ref{e4}) only describes interactions where all three
external legs couple to the heavy quark line, but is valid to all
orders in perturbation theory. Currently results up to order
$\alpha_s^3$ are known \cite{11b}:
\be \label{e5}
{\cal L}_{H_i^0gg}^{(Q)} = - \frac{1}{4} \frac {\alpha_s} { 12 \pi}
F_{\mu \nu a} F^{\mu \nu a} \sum_Q \frac
{c_{iQ}} {M_W} \left[ 1 + \frac {11}{4} \frac {\alpha_s(m_Q)}{\pi} +
\frac {2777 - 201 N_f} {288} \frac {\alpha^2_s(m_Q)} {\pi^2}\right].
\ee
The scale of the overall factor $\alpha_s$ need not be specified,
since the matrix element (\ref{e2}) is scale--independent. However,
the size of the higher order corrections in the square brackets are
determined by $\alpha_s$ taken at the scale $m_Q$, which is the only
energy scale in the problem (note that we are interested in
configurations with essentially vanishing momentum flow through the
diagram). These corrections are therefore larger for $Q=c$ than for
$Q=t$. $N_f$ counts the number of active flavors at scale $m_Q$; we
take $N_f=3,4,5$ for $Q=c,b,t$. \s

The general result eq.~(\ref{e4}) can also be used for squark loop
contributions to the Higgs--gluon coupling. Here results up to ${\cal
O}(\alpha_s^2)$ are known, giving a correction factor $1 + \frac
{25}{6} \frac {\alpha_s(m_{\tilde Q})} {\pi}$ in the effective
Lagrangian, relative to the leading order result listed in
Ref.~\cite{9}. Note that the correction factor is even larger than for
quark loops; however, the overall contributions of squark loops to the
effective $H_i^0gg$ couplings at vanishing external momenta are 
always much smaller \cite{9} than the quark loop contributions. \s

Potentially even larger corrections come from gluino--squark loop
contributions to the $H_i^0 \bar{q} q$ couplings, which are closely
related to $\tgq$ loop corrections to quark masses \cite{12}. These
loop contributions induce a coupling of quarks to the ``wrong'' Higgs 
current eigenstate. In particular, the masses of down--type quarks will now
receive contributions from the vev of the Higgs field with positive
hypercharge. Parametrically, these corrections are therefore of the
order $\delta m_d/m_d \sim \frac{\alpha_s}{\pi} \tanb$. These
corrections can thus become ${\cal O}(1)$ for $\tanb \sim 50$, which
is within the allowed range. \s

The large size of these corrections raises concerns regarding the
reliability of perturbation theory. However, it has recently been
shown \cite{13,14} that the one--loop corrections can be written in
such a way that all higher--order corrections of the form $\left(
\alpha_s/\pi \, \tanb \right)^n$ are re--summed. Note that the
corresponding corrections to the couplings of up--type quarks are much
smaller for $\tanb > 1$; given the uncertainties of the hadronic
matrix elements appearing in the $\tchi N$ scattering amplitude,
corrections of order 10\% or less can safely be ignored. We are thus
only interested in the couplings of down--type quarks $D$ to CP--even
Higgs bosons $h,H$, which can be written as \cite{14}
\beq
s_H^D(\ms) &= \frac {g} {2 M_W} \left[ -m_D^{\rm SM}(\ms)
\cos(\alpha-\beta) + m_D^{\rm MSSM}(\ms) \sin(\alpha-\beta) \tanb
\right] \\
s_h^D(\ms) &= \frac {g} {2 M_W} \left[ m_D^{\rm SM}(\ms)
\sin(\alpha-\beta) + m_D^{\rm MSSM}(\ms) \cos(\alpha-\beta) \tanb
\right]
\label{e6}
\eeq 
where $g$ is the $SU(2)$ gauge coupling, and $\alpha$ is the Higgs
mixing angle in the notation of Ref.~\cite{15}. In eqs.~(\ref{e6})
$m_D^{\rm SM}(\ms)$ stands for the running quark masses, where only
standard QCD corrections are included; their numerical values are
known (with some uncertainties) from experiment. $m_D^{\rm MSSM}(\ms)$
is just the product of the $D$ Yukawa coupling with the vev of the
Higgs field with negative hypercharge; it differs from $m_{D}^{\rm
SM}$ by sparticle loop corrections:
\beq \label{e7}
m_D^{\rm MSSM}(\ms) = m_D^{\rm SM} (\ms) 
- \frac {\alpha_s(\ms)} {3\pi}
m_{\tilde g} \sin (2 \theta_{\tilde D}) \left[ B_0(m_D^2, m_{\tilde g},
m_{\tilde{D}_1}) - B_0(m_D^2, m_{\tilde g}, m_{\tilde{D}_2} ) \right]. 
\nonumber \\
\eeq
Here $\theta_{\tilde D}$ is the $\tilde{D}_L - \tilde{D}_R$ mixing
angle, $m_{\tilde{D}_1}$ and $m_{\tilde{D}_2}$ are the smaller and
larger eigenvalue of the $\tilde{D}$ squark mass matrix, $m_{\tilde
g}$ is the gluino mass, and $\ms$ is a scale of the order of the
squark or gluino mass; we will take the value $\ms = \left( m^2_{\tilde g}
m_{\tilde{D}_1} m_{\tilde{D}_2} \right)^{1/4}$. The first argument of
the Passarino--Veltman two--point function $B_0$ can, to excellent
approximation, be set to zero. In the case of the bottom quark (which
contributes to the $H_i^0 gg$ couplings) we add an additional correction 
from chargino--stop loops, which comes from the large top Yukawa coupling:
\beq \label{e8}
\delta m_b^{\rm MSSM}(\ms) = \frac {g^2} {64 \pi^2} \frac
{m_b^{\rm MSSM}(\ms) m_t(\ms) } {M_W^2 \sin(2\beta) } \mu
\sin(2\theta_{\tilde t})  \left[ B_0(m_b^2,|\mu|,m_{\tilde{t}_1})
 - B_0(m_b^2, |\mu|, m_{\tilde{t}_2}) \right]. \nonumber \\
\eeq
Note that for $\tanb \gg 1$, $\sin(2\theta_{\tilde{D}}) \simeq 2m_D
\mu \tanb / (m^2_{\tilde{D}_2} - m^2_{\tilde{D}_1})$, and
$1/\sin(2\beta) \simeq \tanb/2$; the corrections eqs.~(\ref{e7}) and
(\ref{e8}) are therefore enhanced for large $\tanb$, as advertised. Of
course, if these corrections are set to zero, eqs.~(\ref{e6}) reduce to
the standard tree--level $H_i^0 \bar{q} q$ couplings \cite{15}. \s

It is well known that Yukawa couplings of quarks receive large QCD
corrections, i.e. they show a strong scale dependence. Here we need
these couplings at a relatively low scale, of the order of
the quark mass itself, or in case of light quarks, at a scale not far
from 1 GeV. In contrast, eqs.~(\ref{e6}) should be interpreted as
boundary conditions of the couplings of the effective theory that
emerges when squarks and gluinos are integrated out; these boundary
conditions are thus valid at a scale $\ms$. Below this scale these
couplings run according to standard renormalization group equations
(RGE) \cite{16}:
\be \label{e9}
s^D_{H,h} (Q_1) = s^D_{H,h}(Q_2) \cdot \left[ \frac {\alpha_s(Q_1)}
{\alpha_s(Q_2)} \right]^{12/(33-2N_f)},
\ee
where $N_f$ is again the number of active flavors, and $Q_1$ and $Q_2$
are two energy scales. Since this number changes whenever we cross a
heavy quark threshold, the total effect of the running has to be
computed by applying eq.~(\ref{e9}) repeatedly; e.g.,
\be \label{e10}
s^s_{H,h}(m_c) = s^s_{H,h}(\ms) \cdot  \left[ \frac {\alpha_s(\ms)}
{\alpha_s(m_t)} \right]^{12/21} \cdot  \left[ \frac {\alpha_s(m_t)}
{\alpha_s(m_b)} \right]^{12/23} \cdot  \left[ \frac {\alpha_s(m_b)}
{\alpha_s(m_c)} \right]^{12/25}.
\ee

We have seen in eq.~(\ref{e7}) that the size of the $\tgq$ loop corrections
to $m_D$ is proportional to the amount of $\tilde{D}_L - \tilde{D}_R$
mixing, which in turn is proportional to $m_D$. This raises the
question which $m_D$ to use when computing $\sin(2\theta_{\tilde
D})$. Since the corrections eqs.~(\ref{e7}) and (\ref{e8}) are to be
applied at the high scale $\ms$ it is fairly clear that one should use
a running mass $m_D$ at this high scale. At the one--loop order we cannot
distinguish between the choices $m_{D}^{\rm SM}$ and $m_{D}^{\rm
MSSM}$ here. However, in Ref.~\cite{13} it has been shown that loop
corrections to Higgs boson decay widths into quarks and squarks will
be small only if the tree--level squark mass matrices are written in
terms of $m_{D}^{\rm MSSM}$, i.e. if $m_{D}^{\rm MSSM}$ is used in the
calculation of $\sin(2\theta_{\tilde D})$. This requires an iteration,
which however usually converges quickly. This iteration is equivalent
to the re--summation of higher orders advocated in Ref.~\cite{14}. We
will show below that the difference between using $m_{D}^{\rm SM}$ and
$m_{D}^{\rm MSSM}$ in the calculation of $\sin(2\theta_{\tilde D})$
can indeed be numerically significant. \s

The squark--gluino loop corrections to the mass of $d-$type quarks also 
affect the leading ${\cal O}(m_{\tilde q}^{-2})$ spin--independent
contributions from $d-$type squark exchange. These contributions are
proportional to the quark masses \cite{8,9}, either through the
interference of gauge and Yukawa contributions to the
neutralino--quark--squark couplings (which requires gaugino--higgsino
mixing in the neutralino sector), or through $\tilde{q}_L -
\tilde{q}_R$ mixing. These corrections can again most easily be
understood in terms of an effective $f_q \bar{q} q \bar{\tchi} \tchi$
interaction, where the coefficient $f_q$ is determined by matching to
the full theory at a scale $Q \simeq m_{\tilde q}$ \cite{9}. Since both
the $\tilde{q}_L - \tilde{q}_R$ mixing angle and the quark Yukawa
coupling are determined by $m_{q}^{\rm MSSM}$ rather than by
$m_{q}^{\rm SM}$, the gluino--squark loop correction to the effective
$f_q \bar{q} q \bar{\tchi} \tchi$ interaction can simply be obtained
by multiplying the usual tree--level result \cite{8,9} with
$m_q^{\rm MSSM}(\ms) / m_q^{\rm SM}(\ms)$. Of course, we need these
couplings at a low, hadronic scale. However, the scale dependence of
these couplings is identical to that of the $H \bar{q} q$ couplings,
i.e. is described by eq.~(\ref{e9}). This amounts to a purely
multiplicative renormalization, which leaves the {\em relative} size
of the $\tgq$ loop corrections unchanged. Altogether we thus have
\be \label{e11}
f_D^{(\tilde{D})}|_{\rm improved} = f_D^{(\tilde{D})}|_{\rm tree} \cdot
\frac {m_D^{\rm MSSM} (\ms)} {m_D^{\rm SM}(\ms)},
\ee
where the superscript $(\tilde{D})$ signifies that this expression
only applies to the ${\tilde D}$ squark exchange contribution to the
effective $\bar{D} D \bar{\tchi} \tchi$ interaction\footnote{In fact,
in the ``decoupling limit'', where $\cos(\alpha-\beta) \rightarrow 0$,
the total $\tgq$ loop correction to the $H$ exchange contribution
takes the same simple form as in eq.~(\ref{e11}), while the $h$
exchange contribution is not affected by these corrections.}. \s

The remaining potentially important contributions to the $\tchi N$
scattering amplitude are due to gauge interactions, at least in the
more appealing scenario where the LSP is Bino-- or
photino--like\footnote{We consider this to be more appealing since
one then can obtain an interesting relic density with sparticle masses
in the ``natural'' domain of a few hundred GeV. Higgsino--like LSPs
need masses around a TeV or more to contribute significantly to the
mass density of the Universe.}. These include spin--dependent
contributions due to $Z$ and squark exchange, as well as
spin--independent ${\cal O}(m_{\tilde q}^{-4})$ contributions
\cite{9}. Since to one--loop order electroweak gauge couplings are not
renormalized by strong interactions, we do not expect significant QCD
corrections to these contributions. However, as already stated in the
Introduction, these two contributions are often subdominant.

\section*{3. Results}

We are now ready to present some numerical results. In order to
illustrate the importance of the corrections described in the previous
Section, we show examples for the ratio $R$ of the neutralino
scattering rate on $^{76}$Ge with and without these corrections. If
the small difference between the $\tchi n$ and $\tchi p$ scattering
amplitudes is neglected, $R$ is simply the ratio of the corrected and
uncorrected $\tchi N$ scattering cross sections. \s

In Fig.~(1a,b) we show results for a scenario where all soft SUSY--breaking
contributions to sfermion masses at the weak scale have a common value
$m_0=600$ GeV. Similarly, we took a common value $A_0 = 1.2$ TeV for
all trilinear soft SUSY--breaking parameters at the weak scale. We chose a
relatively small mass of the CP--odd Higgs bosons, $m_A = 240$ GeV
$\simeq 1.6 m_{\tchi}$, in order to obtain LSP detection rates of
interest for present experiments, at least in the region of large
$\tanb$. We assumed the usual unification conditions for gaugino
masses, with an $SU(2)$ gaugino mass $M_2=300$ GeV at the weak scale;
note that now also the gluino mass becomes relevant, see
eq.~(\ref{e7}). We present results for $|\mu| = M_2$ (Fig.~1a) and
$|\mu|=2M_2$ (1b), as a function of $\tanb$. In both cases we show
results for both positive (dashed) and negative (solid)
$\mu$. Finally, the upper (lower) curve with a given pattern has been
obtained using $m_{q}^{\rm MSSM} \ (m_{q}^{\rm SM})$ when calculating
$\sin(2\theta_{\tilde q})$.

\begin{figure}[htb]
\vspace*{-22mm}
\hspace*{1mm}
\mbox{
\epsfig{file=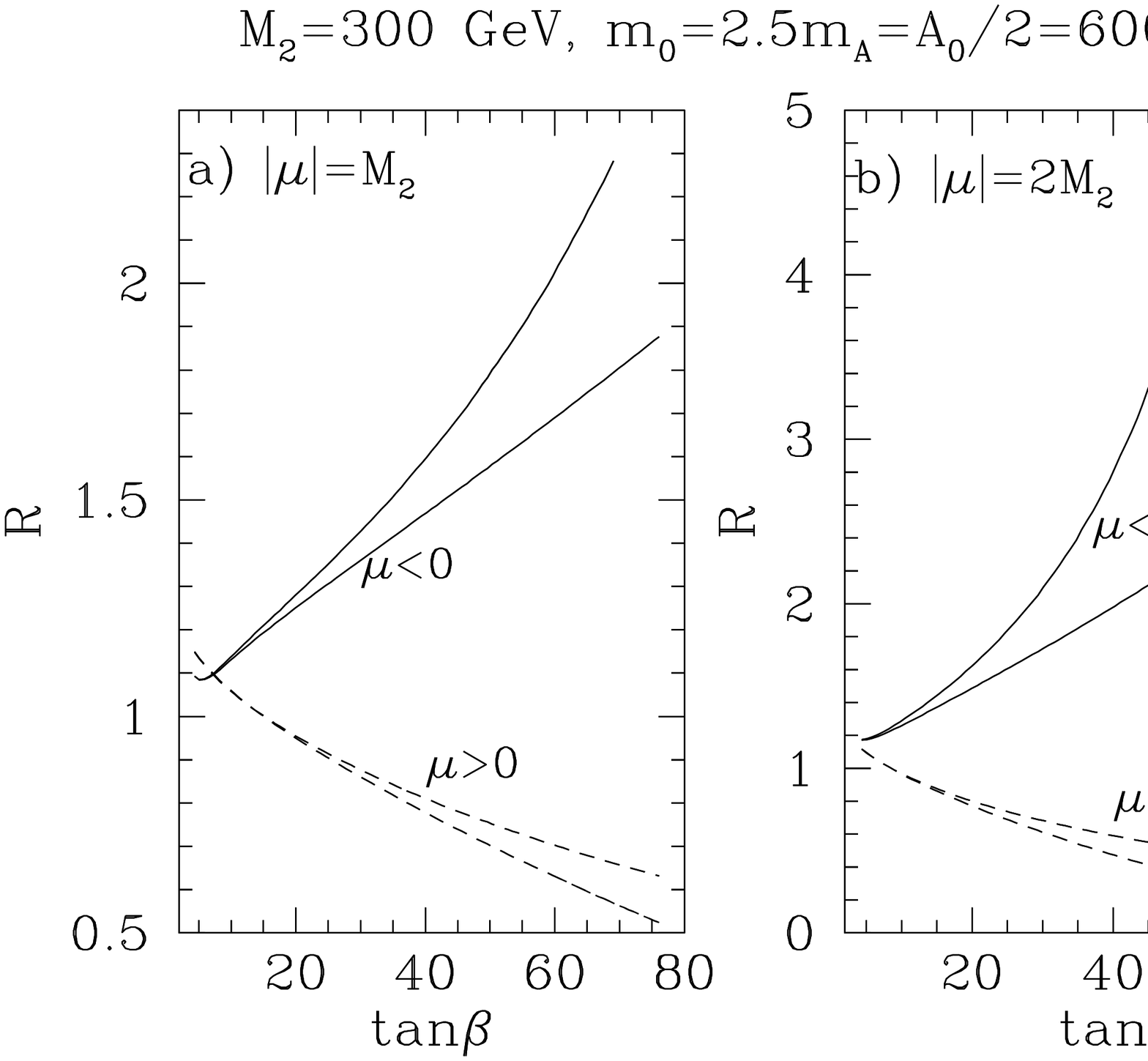,width=0.75\textwidth} }
\vspace*{-22mm}
\caption{The ratio $R$ of corrected to uncorrected $\tchi N$
scattering cross section in a model with universal soft breaking
parameters at the weak scale. The upper (lower) curve of a given
pattern uses the quark mass with (without) sparticle loop corrections
when computing the squark $L$--$R$ mixing angle.}
\end{figure}

For small values of $\tanb$ the total corrections are dominated by the
standard QCD corrections (\ref{e5}) to the $H^0_igg$ couplings. They
increase this contribution to the total $\tchi N$ scattering amplitude
by $\sim 25\%$, which leads to an increase of the counting rate by
$\sim 15 \%$. \s

As anticipated the $\tgq$ loop corrections can be much larger, if
$\tanb \gg 1$. Our sign convention for $\mu$ coincides with that of
Refs.~\cite{9,15}. For $\tanb \gg 1$ the $L-R$ mixing of $d-$type
squarks is essentially proportional to $\mu$; we then find a negative
$\tgq$ loop correction to $m_D$, and hence a positive correction to
the Yukawa coupling for fixed physical quark mass, if $\mu < 0$. Since
it is the Yukawa coupling, rather than the quark mass, which appears
in most spin--independent contributions to the scattering amplitude,
$\mu < 0$ therefore also leads to positive corrections to the
scattering cross section\footnote{The cross--over of the two sets of
curves at small $\tanb$ in Fig.~1a occurs because at small $\tanb$ and
$\mu > 0$ the most important contribution comes from the $h$ boson exchange,
whereas for $\mu<0$ the $H$ boson exchange is always dominant.}. Not
surprisingly, these corrections are considerably larger for larger
$|\mu|$, as can be seen by comparing Fig.~1a with Fig.~1b. \s

Note that we have (somewhat arbitrarily) terminated the curves when
the bottom Yukawa coupling $h_b(\ms) > 1.2$, which roughly corresponds
to its ``fixed point'' value. An even larger $h_b(\ms)$ would lead to
a Landau pole at higher energies, i.e. the Yukawa sector of the theory
would become non--perturbative at energies only slightly above
$\ms$. Since for a fixed value of $\tanb$ the $\tgq$ corrections
increase the Yukawa coupling if $\mu<0$, for this sign of $\mu$ the
upper bound on $\tanb$ becomes significantly stronger; conversely, for
$\mu>0$ larger values of $\tanb$ become accessible once the
corrections are included. Since the predicted counting rate increases
with increasing $\tanb$, the maximal possible $\tchi N$ scattering
cross section for fixed values of the dimensionful input parameters,
which occurs for the biggest allowed value of $\tanb$, would thus not
be affected by the corrections if the relative size of these
corrections were the same for $d, \ s$ and $b$ quarks. \s

Even though we have assumed flavor--independent soft breaking
parameters, this condition is not satisfied. First, while $L$--$R$ mixing
does not change the masses of the $\tilde{d}$ and $\tilde{s}$ squarks
very much\footnote{The difference between the squark masses is in
these cases mostly determined by the different $D-$term contributions
to the masses of $SU(2)$ singlet and doublet squarks.}, it does affect
the $\tilde{b}$ masses significantly. The corrections (\ref{e7}) are
therefore larger, as fraction of the tree--level mass, for $D=b$ than
for $D=d, s$. Moreover, the chargino--stop loop correction (\ref{e8})
{\em only} contributes to the $b-$quark mass. Note that it has the
opposite sign as the $\tgq$ loop correction. For the parameters of
Fig.~1 this is the dominant effect. In fact, in Fig.~1a this top
Yukawa correction is nearly as large as the SUSY--QCD
correction. The reason is that $\tilde{t}_L - \tilde{t}_R$ mixing,
which is relevant in eq.~(\ref{e8}), is proportional to $A_t \ (=A_0$
in our case) which is four times bigger than $|\mu|$ in this
example. As a result, for $\mu<0$ the loop corrections increase the
predicted counting rate at the largest allowed value of $\tanb$ by a
factor of 1.9 (1.4) in Fig.~1a (1b). On the other hand, the increase
of the maximal allowed value of $\tanb$ for $\mu>0$ implies that for
this sign of $\mu$ the maximal rate only decreases by a factor 0.79
(0.74) in Fig.~1a (1b); of course, one also has to keep in mind that
the corrections (\ref{e5}) to the $H_i^0gg$ couplings are always
positive. \smallskip

Note that using $m_{q}^{\rm MSSM}$ rather than $m_{q}^{\rm SM}$ in the
calculation of $\sin(2\theta_{\tilde q})$ always {\em increases} the
prediction of the loop--corrected $\tchi N$ scattering rate. Recall that
a negative correction to the quark mass implies an increase of the
scattering cross section. Since in this case $m_{q}^{\rm MSSM} >
m_{q}^{\rm SM}$, using $m_{q}^{\rm MSSM}$ increases $L$--$R$ mixing and
hence gives a larger, positive correction to the predicted
rate. Conversely, if the corrections to the quark masses for fixed
Yukawa couplings are positive, the correction to the $\tchi N$
scattering amplitude is negative. Since now $m_{q}^{\rm MSSM} <
m_{q}^{\rm SM}$, the use of $m_{q}^{\rm MSSM}$ reduces $L$--$R$ mixing,
and hence lowers the absolute size of the negative corrections to the
scattering rate. \s

The corrections to $h_b$ for fixed physical $m_b$ can have even larger
effects if one assumes universality of soft breaking parameters at
some very high scale, as in the popular mSUGRA models with radiative
$SU(2) \times U(1)_Y$ symmetry breaking \cite{17}. Here one assumes a
common squared soft breaking mass $m_0^2$ not only for sfermions, but
also for Higgs bosons; however, this universality only holds at the
scale of Grand Unification, $M_X \simeq 2 \cdot 10^{16}$ GeV. The
squared soft breaking mass of one of the Higgs bosons is then driven
to negative values by the top Yukawa contribution to the relevant
RGE. One can show \cite{18} that the physical mass of the CP--odd
neutral Higgs boson is then given by\footnote{We ignore small
weak--scale threshold corrections.}
\be \label{e12}
m_A^2 = \frac { m^2_{\tilde \nu} + |\mu|^2 } {\sin^2 \beta}
+ {\cal O}(h_b^2),
\ee
where the corrections from the bottom Yukawa coupling $h_b$ are {\em
negative}. In fact, the upper bound on $\tanb$ often (but not always)
comes from the requirement that $m_A$ should be sufficiently
large. Note that the relevant quantity here is again the bottom Yukawa
coupling, the size of which depends on the size and sign of the $\tgq$
and $\tilde{t} - \tchi^\pm$ loop corrections. The value of $m_A$ is in
turn strongly correlated with the mass $m_H$ of the heavier neutral
CP--even Higgs boson, the exchange of which usually dominates the
spin--independent $\tchi N$ scattering amplitude for $\tanb \gg 1$. In
this kind of model one thus expects a much stronger dependence of the
predicted $\tchi N$ scattering rate on these corrections than in
models where all soft breaking parameters are fixed at the weak scale,
independently of $\tanb$. \s

This is illustrated in Fig.~2, where we took a weak--scale $SU(2)$
gaugino mass of 200 GeV, and GUT--scale boundary conditions $m_0 = A_0
= 250$ GeV. Note that in this model the absolute size of $\mu$ is a
derived quantity, but its sign can still be chosen freely; dashed
(solid) lines are again for positive (negative) $\mu$. The two inner
curves have been obtained by using $m_{q}^{\rm SM}$ when calculating
$L$--$R$ squark mixing, as well as $h_b$ as it appears in the RGE. In
this case the correction ``only'' amounts to a factor of 2 at most,
similar to the case shown in Fig.~1a. \s

\begin{figure}[htb]
\vspace*{-16mm}
\hspace*{15mm}
\mbox{
\epsfig{file=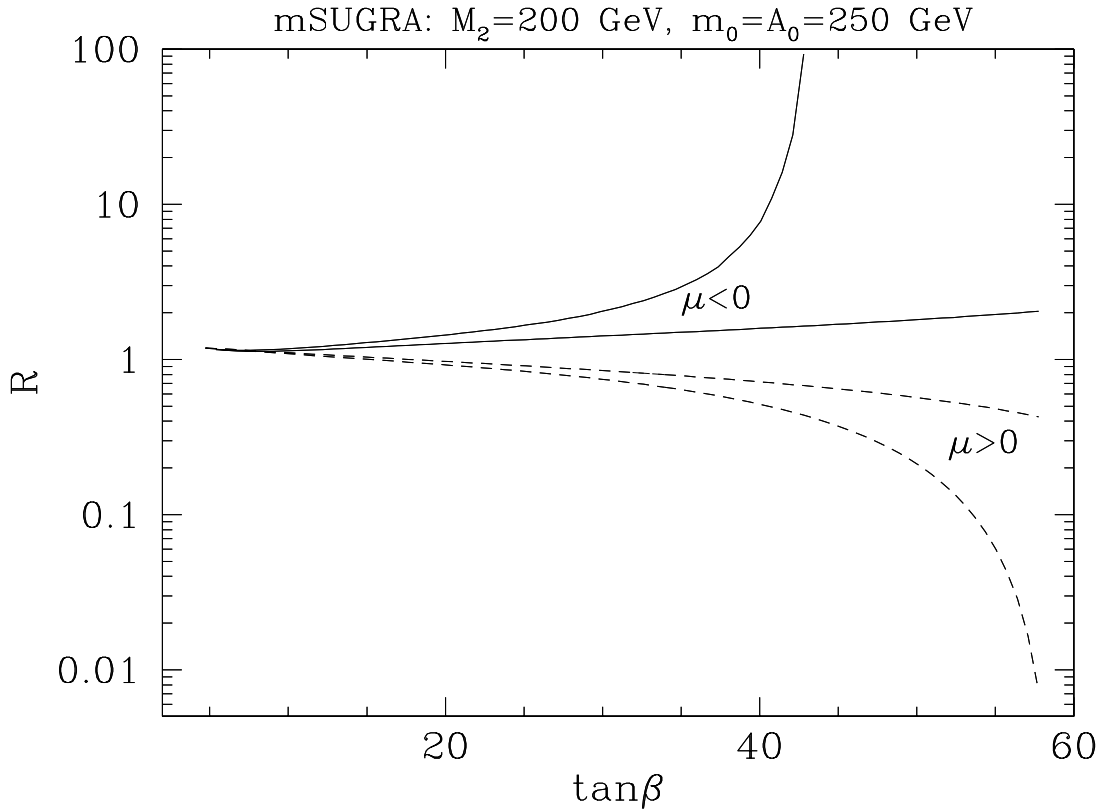,width=0.7\textwidth} }
\vspace*{-15mm}
\caption{The ratio $R$ of corrected to uncorrected $\tchi N$
scattering cross section in an mSUGRA model. The outer (inner) pair of
curves uses the quark mass with (without) sparticle loop corrections
when computing the squark $L-R$ mixing angle as well as the $b$--quark
Yukawa contribution to the RGE.}
\end{figure}

The outer curves in Fig.~2 have been obtained by using $m_{q}^{\rm
MSSM}$ both in the calculation of $\sin(2\theta_{\tilde q})$ and in
the calculation of the bottom quark Yukawa coupling in the RGE. Mostly due to
the change of $m_A$ discussed above, we now find correction factors as
large as 100 for fixed values of $\tanb$. Note that in the given
example for $\mu<0$ the upper bound on $\tanb$ is determined by the
lower bound on $m_A$; we have required $m_A + m_h > 180$ (150) GeV if
$\cos^2(\alpha-\beta) \simeq 1 \ (\geq 0.2)$, to cope with the experimental 
constraints. As a result the predicted
counting rate at the maximal allowed value of $\tanb$ changes very
little when the loop corrections are included.\footnote{Note that we
always include the $\tgq$ loop corrections to the mass matrix of
neutral Higgs bosons \cite{19}. The dominant contribution here usually
comes from $\tilde{t} - \tilde{g}$ loop corrections to $m_t$, but for
large $\tanb$ and small $m_A$ values, $\tilde{b}-\tilde{g}$ loop corrections
to $m_b$ can also have a significant effect on the Higgs mixing angle
$\alpha$.} On the other hand, for $\mu>0$ the upper bound on $\tanb$
after the inclusion of sparticle loop corrections comes from the
requirement that the lighter $\tilde{\tau}$ mass eigenstate should not
be lighter than the lightest neutralino. In this case the corrections
therefore reduce the maximal possible $b$ (and $s$ and $d$) Yukawa
couplings, leading to a reduction of the predicted $\tchi N$
scattering rate at the maximal allowed value of $\tanb$ by a factor close
to 100. 

\section*{4. Summary and Conclusions}

In this paper we discussed two kinds of (potentially) large QCD
corrections to the $\tchi N$ scattering amplitude. Standard QCD
corrections increase the effective coupling of CP--even Higgs bosons
to two gluons, which proceeds through a loop of heavy quarks, by $\sim
25\%$. Supersymmetric QCD corrections to the quark masses, or to their
Yukawa couplings for fixed physical quark masses, can have even larger
effect if $\tanb \gg 1$: we found that an enhancement or suppression
of the predicted $\tchi N$ scattering cross section by a factor of 2
is quite easily possible for a fixed set of soft breaking parameters
and fixed $\tanb$. Since the correction can also change the maximal
allowed value of $\tanb$, the change of the maximal possible counting
rate, which occurs at the largest allowed value of $\tanb$, is usually
smaller, but can still be significant. \s

Much bigger effects from $\tgq$ loop corrections are possible if the
size of the $b$ Yukawa coupling affects the particle spectrum at the
weak scale, as in mSUGRA models. The dominant effect here comes from
the anti--correlation between the $b$ Yukawa coupling and the masses
of the heavier Higgs bosons. Since for large $\tanb$ the scattering
cross section roughly scales like the inverse fourth power of the mass
of the heavier neutral CP--even Higgs boson, the sparticle
loop--induced change of the $b$ Yukawa coupling can change the
predicted neutralino detection rate by up to two orders of magnitude,
if $\tanb$ is large.\footnote{This indirect effect is included in the
recent analysis by Accomando et al. \cite{20}, but its importance is
not discussed explicitly there.} Note that $\tanb \gg 1$ is required
in this model to obtain scattering rates that are sufficiently large
to be probed in present or near--future experiments. \s

One consequence of these sparticle loop corrections is a strong
dependence of the predicted $\tchi N$ scattering rate on the sign of
$\mu$ even for $\tanb \gg 1$. It is well known that for small and
moderate $\tanb$, positive values of $\mu$ lead to more
gaugino--higgsino mixing in the neutralino eigenstate, and hence to
larger couplings of the neutralino to Higgs bosons, than do negative
values of $\mu$. This tree--level dependence of the scattering rate on
the sign of $\mu$ disappears at large $\tanb$. However, there the
$\tgq$ loop corrections become important. Note that now $\mu<0$ gives
larger counting rate (for fixed $|\mu|$) than $\mu > 0$. These
corrections can thus qualitatively change the dependence of the
predicted relic neutralino detection rate on the parameters of the
supersymmetric model.

\bigskip

\noindent {\bf Acknowledgements:} 
The work of A.D. was partially supported by the French
GDR--Supersym\'etrie. The work of M.D. was supported in part by the
``Sonderforschungsbereich 375--95 f\"ur Astro--Teilchenphysik'' der
Deutschen Forschungsgemeinschaft.

\end{document}